# Single-Molecule Surface-Enhanced Raman Scattering Spectrum of Non-Resonant Aromatic Amine Showing Raman Forbidden Bands

*Yuko S. Yamamoto,[a,b]* Yuya Kayano,[a] Yukihiro Ozaki,[c] Zhenglong Zhang,[d] Tomomi Kozu,[e] Tamitake Itoh,[f] and Shunsuke Nakanishi[a]* a Department of Advanced Materials Science, Faculty of Engineering, Kagawa University, Kagawa 2217-20, Takamatsu, Japan.

b Research Fellow of the Japan Society for the Promotion of Science (JSPS), Tokyo 102-0083, Chiyoda, Japan.

c Department of Chemistry, School of Science and Technology, Kwansei Gakuin University, Sanda, Hyogo 669-1337, Japan.

d Leibniz Institute of Photonic Technology, Albert-Einstein-Str. 9, 07745 Jena, Germany.

e Field of Systems Design Engineering, Department of Integrated Engineering Science, Akita University, Akita 010-8502, Tegatagakuen, Japan.

f National Institute of Advanced Industrial Science and Technology (AIST), Kagawa 761-0395, Takamatsu, Japan

*E-mail: yamayulab@gmail.com

We present the experimentally obtained single-molecule (SM) surface-enhanced Raman scattering (SERS) spectrum of 4-aminobenzenethiol (4-ABT), also known as para-aminothiophenol (PATP). Measured at a 4-ABT concentration of 8 × $10^{-10}$ M, the spectra show Raman forbidden modes. The SM-SERS spectrum of 4-ABT obtained using a non-resonant visible laser is different from the previously reported SERS spectra of 4-ABT, and could not be reconstructed using quantum-mechanical calculations. Careful classical assignments (not based on quantum-mechanical calculations) are reported, and indicate that differences in the reported spectra of 4-ABT are mainly due to the appearance of Raman forbidden bands. The presence of Raman forbidden bands can be explained by the charge-transfer (CT) effect of 4-ABT adsorbed on the silver nanostructures, indicating a breakdown of Raman selection rules at the SERS hotspot.

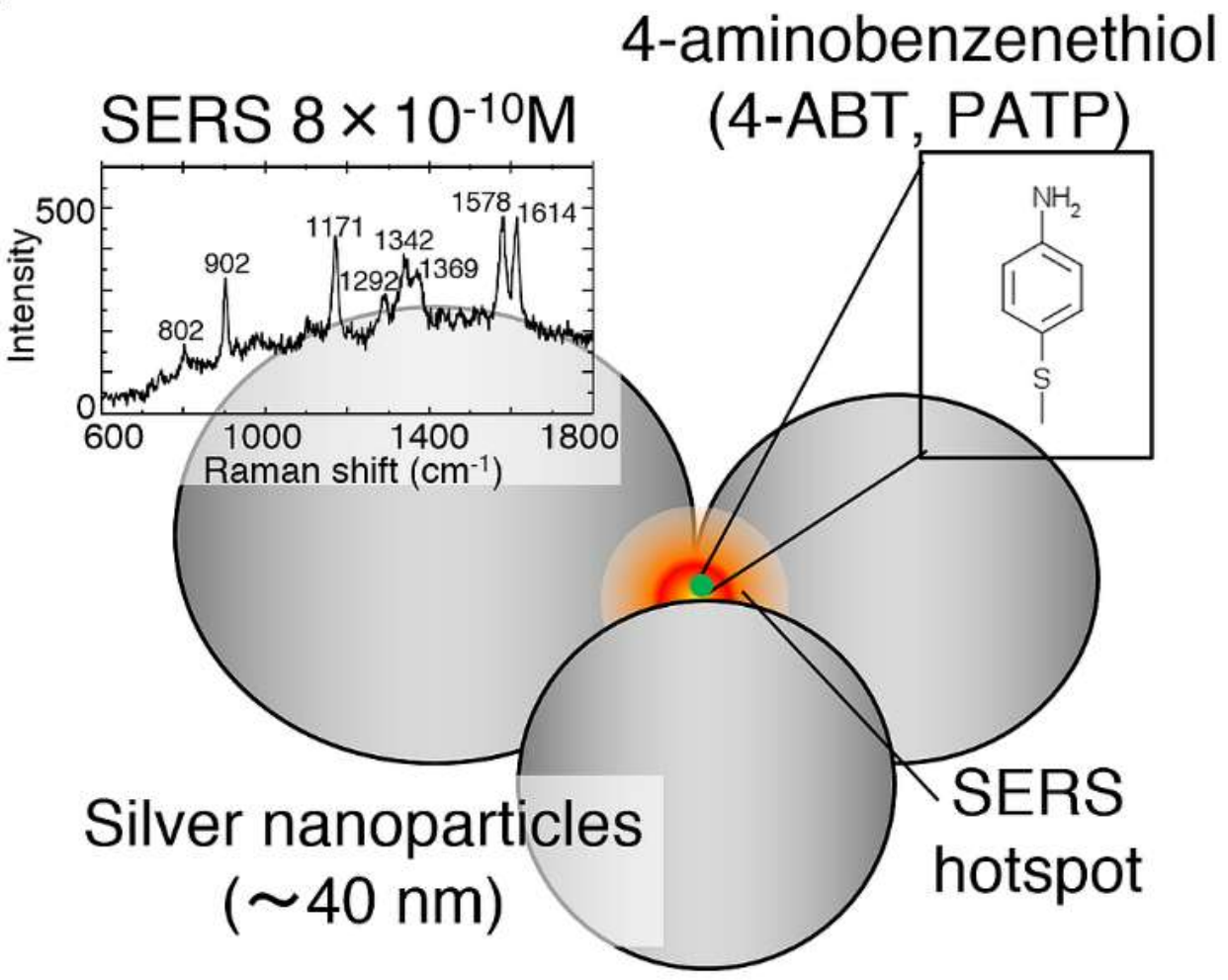

Surface-enhanced Raman scattering (SERS) is widely known as an ultrasensitive vibrational spectroscopic technique with sensitivity at the single-molecule level.[1] The SERS cross-section of the target molecule is largely enhanced when the molecule is located on the surface of silver or gold nanostructures, which exhibit huge plasmon resonance. Enhancement mechanisms of SERS have been classified into two categories, electromagnetic mechanism (EM) and chemical effects (CE).[2] The EM mechanism in SERS, which is directly induced by plasmon resonances of metal nanostructures, has long been discussed[3] and experimentally confirmed using single-particle spectroscopy.[4] While, CE effect in SERS still remains unclear. CE is mainly classified into two types of resonances:[5,6] (a) classical resonance effect of Raman scattering, and (b) charge-transfer (CT) resonance due to nanostructure–molecule CT transitions.

4-Aminobenzenethiol (4-ABT), also known as para-aminothiophenol (PATP), has been used as an important probe of CT effect in SERS. This is because 4-ABT forms a CT complex to a silver or gold surface via its thiol residue, resulting in a very different SERS spectrum from its normal Raman spectrum.[7,8] Originally, the CT effect of 4-ABT was explained by Herzberg-Teller (vibronic) coupling between the Fermi level of the metal nanostructure and the LUMO of the 4-ABT attached on the metal nanostructure. This discussion was continued based on the SERS spectrum measured of self-assembled monolayers of 4-ABT.[9-11] However, in recent years, it was found during SERS measurements of 4-ABT self-assembled monolayers that 4-ABT forms its dimer, 4,4'-dimercaptoazobenzene (DMAB).[12] This means that the original discussion of CT effects by means of 4-ABT was based on the SERS spectrum of DMAB and not 4-ABT. This forces a modification of the model. Moreover, the actual SERS spectrum of a single 4-ABT molecule is still missing. In this letter, we present the first single-molecule SERS (SM-SERS) spectrum of 4-ABT. To achieve SM-SERS, silver colloidal nanoparticle (NP) aggregates were used following the method first reported in 1997.[13,14] This method creates SERS hotspots where the Raman cross-section is enlarged by a factor of ~$10^{10}$, allowing SM-SERS measurements of 4-ABT.[15] We finally achieved the SERS measurements of 4-ABT at a concentration of $8 \times 10^{-10}$ M. This is the first report of the experimentally obtained SERS spectrum of 4-ABT at such low concentrations and the present results provide new insight into the CT effect in SERS.

Figure 1 shows the representative SERS and normal Raman (NR) spectra of 4-ABT. To obtain the SM-SERS measurements, the concentration of 4-ABT was reduced from $8 \times 10^{-6}$ to $8 \times 10^{-10}$ M. At concentrations of $8 \times 10^{-6}$ to $8 \times 10^{-8}$ M, the experimentally obtained SERS spectra are identical to those obtained previously.[7-12,17,18-24,26] Therefore, we concluded that these spectra arose due to the presence of dimers of 4-ABT, i.e., DMAB. In contrast, the SERS spectral shapes at lower concentrations ($8 \times 10^{-9}$ M and $8 \times 10^{-10}$ M) differ from those measured between $8 \times 10^{-6}$ and $8 \times 10^{-8}$ M. The concentrations of $8 \times 10^{-9}$ and $8 \times 10^{-10}$ M can be estimated as the concentrations required for SM level SERS measurements.[13,14,25] Consequently, we expected these SERS spectra to be identical to the NRS spectrum of 4-ABT.[7,10,17,19-23,27,29,32] However, the SERS spectral shapes are significantly different from those of the experimentally obtained normal Raman scattering (NRS) spectrum of 4-ABT.

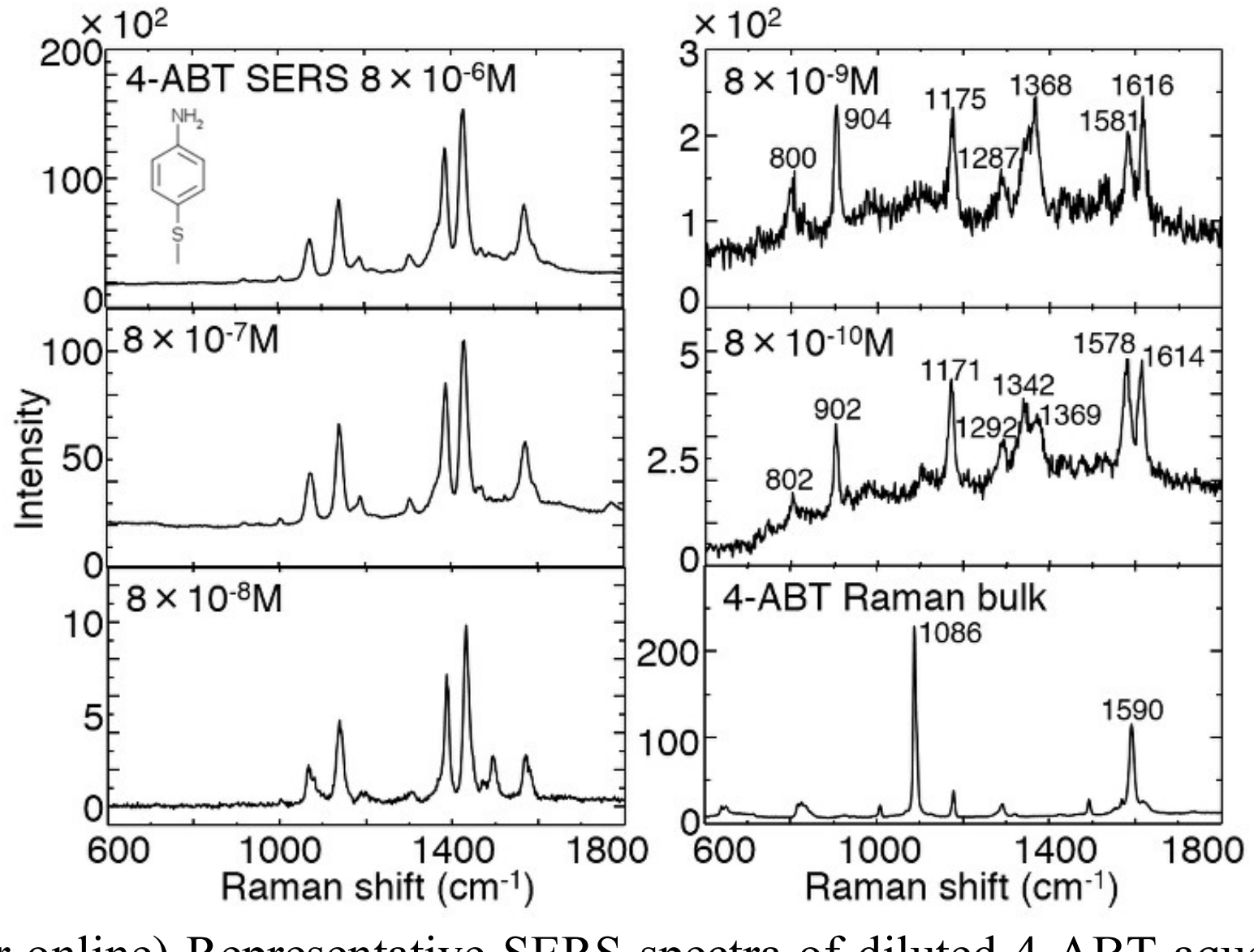

Figure 1. (Color online) Representative SERS spectra of diluted 4-ABT aqueous solutions and NRS spectrum of neat 4-ABT. The concentration of each 4-ABT solution is indicated in the panel.

Here, we evaluate whether the most dilute 4-ABT concentration used ($8 \times 10^{-10}$ M) under the present measurement conditions truly represent SM-SERS conditions. We used Nie's method of estimation,[14] which is known as one of the first SM-SERS measurements, to find the number of 4-ABT molecules attached to each Ag NP. The final concentration of the Ag NP colloid used in this study was estimated to be $1 \times 10^{-10}$ M by comparison with Mie scattering calculations. In the most dilute 4-ABT solution, the concentration of 4-ABT was $8 \times 10^{-10}$ M; therefore, each Ag NP has an average of 8 molecules, assuming that all molecules in the solution are attached to the surfaces of Ag NPs. We estimate that fewer than 80% of available 4-ABT molecules were adsorbed on the surfaces of the Ag NP after incubation for 1 h; that is, on average, six 4-ABT molecules are adsorbed to the Ag NP surface. The average distance, d, between adjacent molecules can be estimated by the following equation:[25]

$$d=2R\sqrt{(\pi/n)} \qquad (1)$$

where R is the average radius of the Ag NPs (20 nm) and n is the number of molecules absorbed on each Ag NP (six). Consequently, the calculated average distance, d (29 nm) is more than 29 times larger than the size of a 4-ABT molecule (0.98 nm at maximum length). Therefore, we concluded that, at the most dilute concentration of 4-ABT ($8 \times 10^{-10}$ M) 4-ABT is present as monomers and does not form dimers on the surfaces of the Ag NPs. This satisfies the measurement conditions for SM-SERS. Moreover, the SERS measurement at the higher concentrations of $8 \times 10^{-9}$ M yields an almost identical spectrum to that at $8 \times 10^{-10}$ M; therefore, we further concluded that, the SM-SERS measurement conditions are also met at a concentration of $8 \times 10^{-9}$ M.

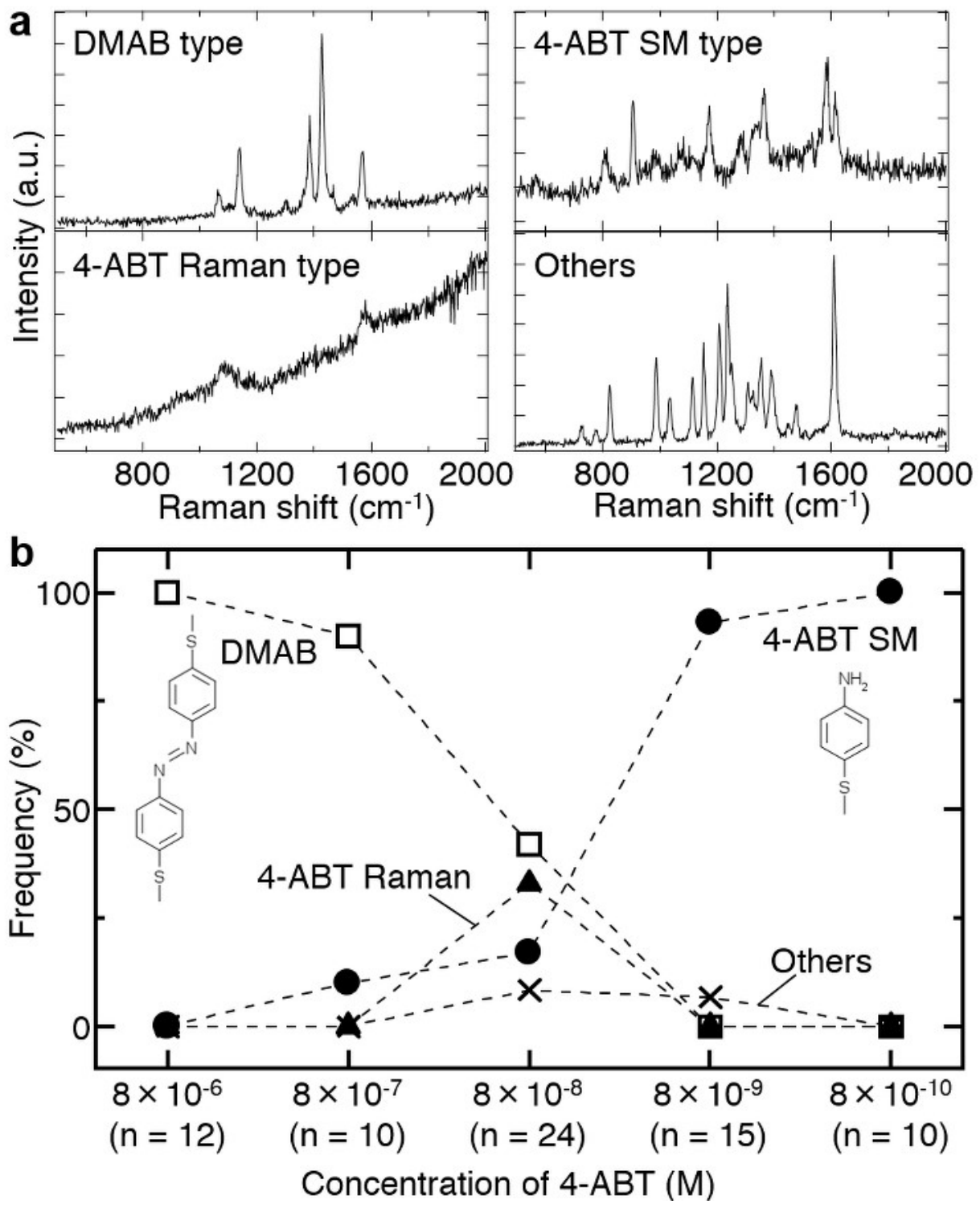

Figure 2. (a) SERS spectral variation of 4-ABT at the concentration of $8 \times 10^{-6}$ to $8 \times 10^{-10}$ M. (b) Appearance probabilities of four types of SERS spectrum as a function of 4-ABT concentration.

For further evidence of SM-SERS, we discuss the SERS spectral variation at each concentration. Figure 2a shows the spectral variation in the SERS spectra of 4-ABT at $8 \times 10^{-6}$ and $8 \times 10^{-10}$ M. At these concentrations, three main types of SERS spectra were detected, as well as small amounts of other types of spectra. However, the appearance probability of each SERS spectrum is different at each concentration. The appearance probabilities of the four types of SERS spectrum are summarized in Fig. 2b. As shown in Fig. 2a and 2b, the DMAB-type SERS spectrum is the main spectrum at concentrations from $8 \times 10^{-6}$ to $8 \times 10^{-8}$ M. At a concentration of $8 \times 10^{-8}$ M, a Raman-type SERS spectrum of 4-ABT is also detected. At lower concentrations, namely $8 \times 10^{-9}$ and $8 \times 10^{-10}$ M, the SM-SERS spectrum of 4-ABT predominates. On decreasing the concentration of 4-ABT, the transition from a DMAB-type SERS spectrum to an SM-SERS spectrum of 4-ABT is clearly seen, again indicating that, at these low concentrations, SM-SERS conditions are met. Previous investigations have not reported this type of spectral variation. Therefore, careful assignment of the SERS peaks is required.

Table 1 shows the peak assignments of the 4-ABT NRS, 4-ABT SM-SERS, and DMAB SERS spectra shown in Fig. 2a. Note that the DFT-calculated 4-ABT NRS spectrum does not fit that experimentally obtained for 4-ABT (see Supporting Information). Moreover, the DFT calculations of 4-ABT interacting with metal clusters, imitating the SERS measurement conditions, yield a spectrum that is similar to the experimentally obtained 4-ABT NRS spectrum, and not the SERS spectrum.[17a,27] These facts imply that reproduction of the 4-ABT NRS and SERS spectra by previous DFT calculation methods have been unsatisfactory. Therefore, in this study, we alternatively employ classical assignment procedures[28] and do not use DFT calculations to assign the peaks of the experimentally obtained 4-ABT NRS and SERS spectra. The assignments reported by Osawa *et al.*[7] were additionally used to assign the NRS spectrum of 4-ABT since their assignments were not obtained from DFT calculations. In contrast, calculation of the DMAB SERS spectrum using DFT yields a predicted spectrum that matches that obtained experimentally.[17a,26,29] Therefore, the DFT assignments given in Ref. 26 are quoted in Table 1 for comparison. On comparing, the SM-SERS spectrum with the NRS spectrum of 4-ABT, it is clear that some of the peaks in the NRS spectrum of 4-ABT (peaks at 634, 1011, 1086, 1266, and 1490 $cm^{-1}$) are absent in the SM-SERS spectrum. In addition, several new peaks (802, 1292, 1342, and 1369 $cm^{-1}$) are present in the SM-SERS spectrum. Note that the 4-ABT NRS spectrum at 820 $cm^{-1}$ and the 4-ABT SM-SERS band at 802 $cm^{-1}$ may arise from the same vibration, assigned to CH wagging(Table 1).

**Table 1.** Experimentally obtained NRS and SM-SERS frequencies of 4-ABT in the range of 1800 – 600 $cm^{-1}$. Experimentally-obtained DMAB SERS frequencies are also shown.

| 4-ABT Normal Raman | 4-ABT Single-molecule SERS ($8\times10^{-10}$ M) | Assignments[7, 28] | DMAB SERS | Assignments[26] |
|---|---|---|---|---|
| 1620 vw | | δNH [7] | | |
| 1590 s | 1614, 1578 | νCC (8a)($a_1$) [7] | 1578 m | νCC (47)($a_g$) |
| 1490 w | | νCC+δCH (19a)($a_1$) [7] | | |
| | | | 1431 s | νNN(31)+βCH(29) ($a_g$) |
| | | | 1387 s | νNN(32)+βCN(30) ($a_g$) |
| | 1369 | νArN (IR) [28] | | |
| | 1342 | νArN (IR) [28] | | |
| | | | 1306 w | νCC(80) ($a_g$) |
| | 1292 | νArN (IR) [28] | | |
| 1266 w | | - | | |
| | | | 1194 w | νCN(26)+βCH(41) ($a_g$) |
| 1173 m | 1171 | δCH (9a)($a_1$) [7] | | |
| | | | 1144 s | νCN(23)+βCH(49) ($a_g$) |
| 1086 vs | | νCS (7a)($a_1$) [7] | | |
| | | | 1075 m | νCC(47)+νCS(20)($a_g$) |
| 1011 w | | γCC+γCCC(18a)($a_1$) [7] | | |
| | 902 | δSH (IR) [28] | | |
| 820 w | | πCH (11)($b_1$) [7] | | |
| | 802 | π [2 Adjacent H] [28] | | |
| 634 m | | γCCC (12)($a_1$) [7] | | |

ν; stretching, β; bending, π; wagging, γ and δ; bending, vs; very strong, s;strong, m; medium, s; small, IR: possible assignment from Infrared in ref. 30 due to charge-transfer contribution.

Notably, the triplet peaks appearing at 1292, 1342, and 1369 $cm^{-1}$ in the SM-SERS spectrum can be assigned to the stretching modes of the primary aromatic amines with nitrogen directly attached to the benzene ring; these are Raman forbidden peaks.[28] Moreover, the peak at 902 $cm^{-1}$ in the same spectrum can also be assigned to a Raman forbidden peak that arises due to the deformation of an S-H bond.[28] In principle, such peaks are usually infrared (IR) active and are not detectable in the SERS spectra. However, if a detection condition causes a rigorous resonance Raman condition, the Raman selection rule changes due to vibronic coupling, and forbidden Raman peaks become detectable..[6,8,30] The absorption spectrum of 4-ABT (see Supporting Information) shows that the resonance Raman condition lies in the ultraviolet region (< 350 nm). Thus, the appearance of IR peaks indicates that the resonance Raman condition is met by the formation of molecule-to-metal charge transfer (CT) complexes[2,5,31,32] on the metal surface. On forming CT complexes, the energy required for electronic transitions of the 4-ABT molecule decreases and the excitation laser line meets the resonance Raman condition. CT resonance is also consistent with the present SM-SERS detection because the Raman cross-section of 4-ABT, $\sim10^{-29}$ $cm^2$, cannot reach the value required for SM detection ($\sim10^{-16}$ $cm^2$) under the current experimental conditions, even given the maximum EM enhancement factor ($< 10^{10}$) (see Supporting Information). Therefore, the additional enhancement factor ($> 10^3$), due to CT resonance,[6,32,33] may enable the SM-SERS detection of 4-ABT molecules observed in the present study. Indeed, the presence of 4-ABT Raman-type spectra in Fig. 2a, which are free from CT resonance and are always much weaker in intensity than both the 4-ABT SM-SERS and DMAB SERS spectra, also supports the existence of CT resonance in the 4-ABT SM-SERS measurements. These differences in the spectra also indicate that the previously reported experimentally obtained 4-ABT SERS spectra,[27] which are similar to the experimentally obtained NRS spectrum of 4-ABT, were the sum of SERS signals from many 4-ABT molecules that had not formed CT complexes at the metal surface.

In summary, we have demonstrated SM-SERS measurements of a single 4-ABT molecule obtained by diluting an aqueous solution of 4-ABT from $8 \times 10^{-6}$ M to $8 \times 10^{-10}$ M. This experimentally obtained SM-SERS spectrum of 4-ABT is novel and differs from the experimentally obtained SERS spectrum of DMAB, the NRS spectrum of 4-ABT, and the results of DFT calculations for both DMAB and 4-ABT. The concentration dependence of the experimentally obtained SERS spectrum of 4-ABT is shown by a transition from a DMAB-type SERS to an SM-SERS of 4-ABT on decreasing the concentration of 4-ABT. The SM-SERS spectrum of 4-ABT shows Raman forbidden peaks that may arise due to the CT resonance Raman effect,[2,5,31,32] which is an additional factor to EM enhancement[6,32,33] that allows realization of the SM-SERS measurement under the conditions used in the present study.

**Experimental Section**

To achieve SM-SERS detection of single 4-ABT molecules, a colloidal silver solution was used, as in typical SERS measurements.[13,14] See supporting information for details. Briefly, a Ag NP colloidal solution ($1 \times 10^{-10}$ M) was prepared by the reduction of $AgNO_3$ with sodium citrate according to the method of Lee and Meisel.[16] 4-ABT mother solution in methanol was diluted with distilled water and then the Ag NP colloidal solution and the NaCl aqueous solution were sequentially added to obtain Ag NP nanoaggregates which scatter SERS signal of 4-ABT. The SERS spectrum was measured using a customized SERS microscope system.

We thank Mr. Katsuhiro Saito (Tokyo Instruments Co., Ltd) for his support with high-sensitivity instrumentation. This work was supported by a JSPS KAKENHI Grant-in-Aid for Scientific Research (C) number 20510111, (B) number 26286066, and WAKATE (B) number 26810013. This work was also supported by a Research Fellowship RPD of the Japan Society for the Promotion of Science (JSPS) Number 26–40216.

## Supporting information

### S1. Experimentally-obtained normal Raman, calculated Raman, and absorption spectra of 4-aminobenzenethiol

Figures S1a and S1b show the experimentally obtained normal Raman spectrum of 4-aminobenzenethiol (4-ABT, *para*-aminothiophenol, PATP) in the neat state and density functional theory (DFT) calculations predicting the normal Raman spectra (NRS) of 4-ABT. For the experimental normal Raman measurement, a 532 nm continuous wave (CW) diode laser was used as a light source. The density functional theory (DFT) calculations including structure optimization and Raman frequency computation were carried out by using Gaussian 09. The hybrid exchange correlation functional B3LYP was used to predict the NRS spectrum of 4-ABT. The 6-31G(d) basis sets for C, N, S, and H atoms were used. Note that the calculated NRS spectrum (Fig. S1b) is almost identical to the calculated Raman spectrum reported in Ref. S1. Figure S1c shows the experimentally obtained absorption spectrum of 4-ABT in methanol solution ($10^{-4}$ M). A green line indicates the incident laser line (532 nm) used in the present normal Raman and SERS measurements.

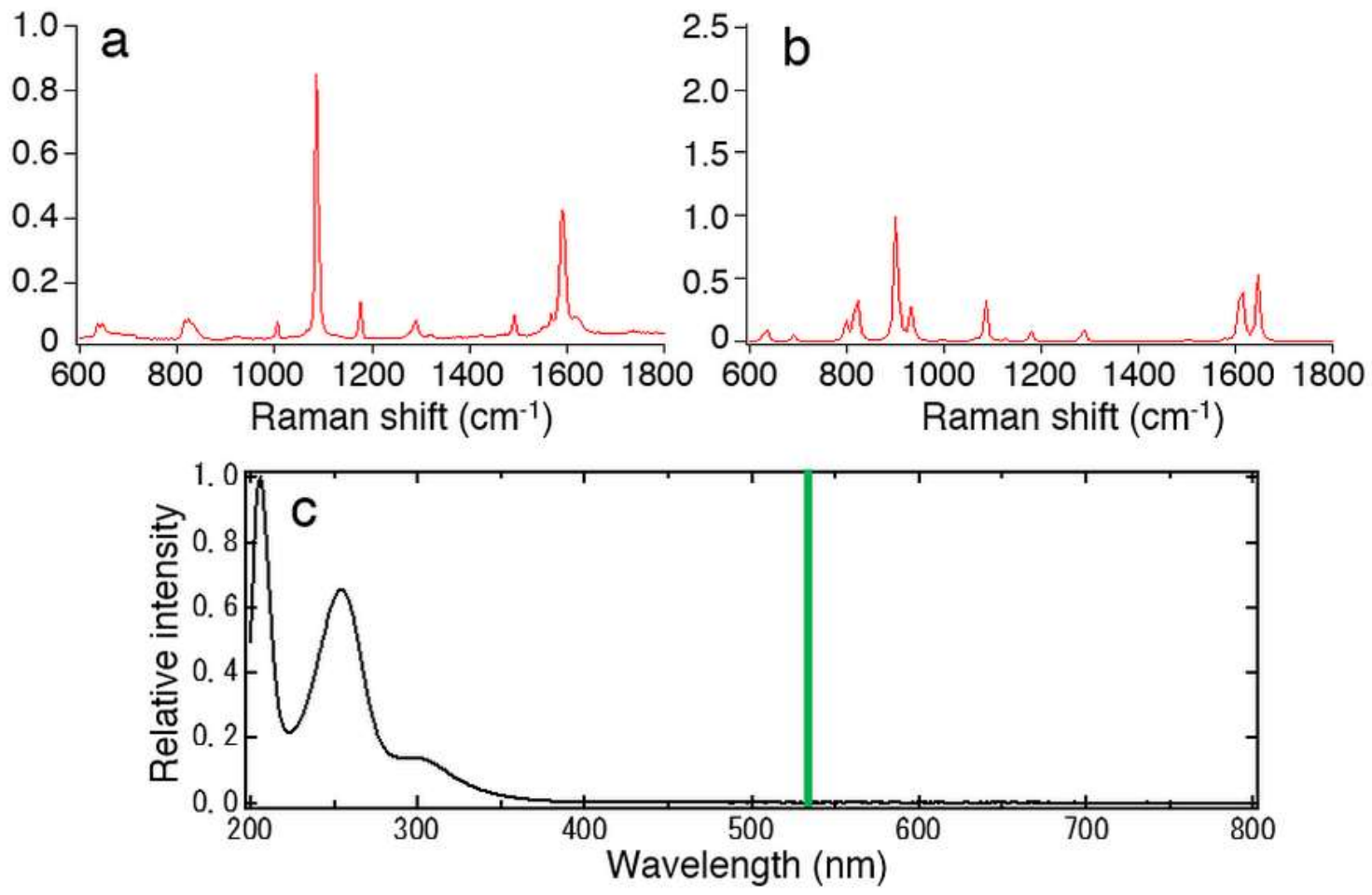

Figure S1. (a) Experimentally obtained NRS spectrum of 4-ABT in neat state. (b) DFT prediction of NRS spectrum of 4-ABT. (c) Experimentally obtained absorption spectrum of 4-ABT in methanol solution ($10^{-4}$ M).

**S2. Experimental**

To achieve SM-SERS detection of single 4-ABT molecules, a colloidal silver solution was used, as in typical SERS measurements.[RS2,3] Briefly, a Ag NP colloidal solution ($1 \times 10^{-10}$ M) was prepared by the reduction of $AgNO_3$ (Wako Chemicals, Japan) with sodium citrate (Wako Chemicals, Japan) according to the method of Lee and Meisel.[RS4] The average diameter of the AgNP colloids was 40 nm. To carry out the SM-SERS measurements, mother solution of 4-ABT (Wako chemicals, Japan) ($10^{-4}$ M in methanol) was diluted with distilled water. Then, the Ag NP colloidal solution was added to the 4-ABT dilution. After waiting from between 30 min to 1 h, a 1 M NaCl aqueous solution was added to the mixture to obtain AgNP nanoaggregates (see Supporting Information). The final concentration of NaCl was 10–20 mM. The concentration of the 4-ABT dilutions from $8 \times 10^{-6}$ to $8 \times 10^{-10}$ M, the final concentration being based that used in the standard SM-SERS method (~$10^{-10}$ M). Then, the AgNP nanoaggregate solution was sandwiched between a ranged pre-cleaned slide glass (S7224, Matsunami, Japan) and a cover slip (18 mm × 18 mm, thickness no. 1, Matsunami, Japan). The surface morphologies of the silver nanoaggregates were evaluated by scanning electron microscopy (SEM; JSM-6700F, JEOL, operating at 15 kV) after drying these nanoaggregates in a vacuum chamber. The length of a single 4-ABT molecule was determined using Winmostar V6.013 software (X-Ability Co., Ltd., Japan) based on the van der Waals radii.

The SERS spectrum was measured using a customized SERS microscope system, which is similar to other experimental setups.[RS5] Figure S2 shows the customized inverted microscope system used for SERS measurements. A green laser beam (532 nm, 3.5 W/cm$^2$) for SERS excitation is focused on the sample mounted on the stage by using a lens besides the inverted microscope (IX-71, Olympus, Japan). The sample is illuminated with white light from a 50-W halogen lamp through a dark-field condenser to detect elastically scattered light. Elastic and SERS scattering lights from the identical nanoaggregate are collected through an objective lens and sent to a polychromator (SR303i-AT, Andor Technology Ltd., UK) and a CCD detector (DV420A-0E, Andor, Andor Technology Ltd., UK). Note that spectra of SERS intensity $I_{SERS}(\lambda_L,\lambda)$ (photocounts) can be converted into the SERS cross section spectra $\sigma_{SERS}(\lambda_L,\lambda)$ ($cm^2$) using a 80-nm gold nanosphere, whose scattering intensity and scattering cross section are known.[RS6] Note that the conversion factor is $1.3 \times 10^{-18}$ (photocounts/$cm^2$) on the current spectroscopic equipment.

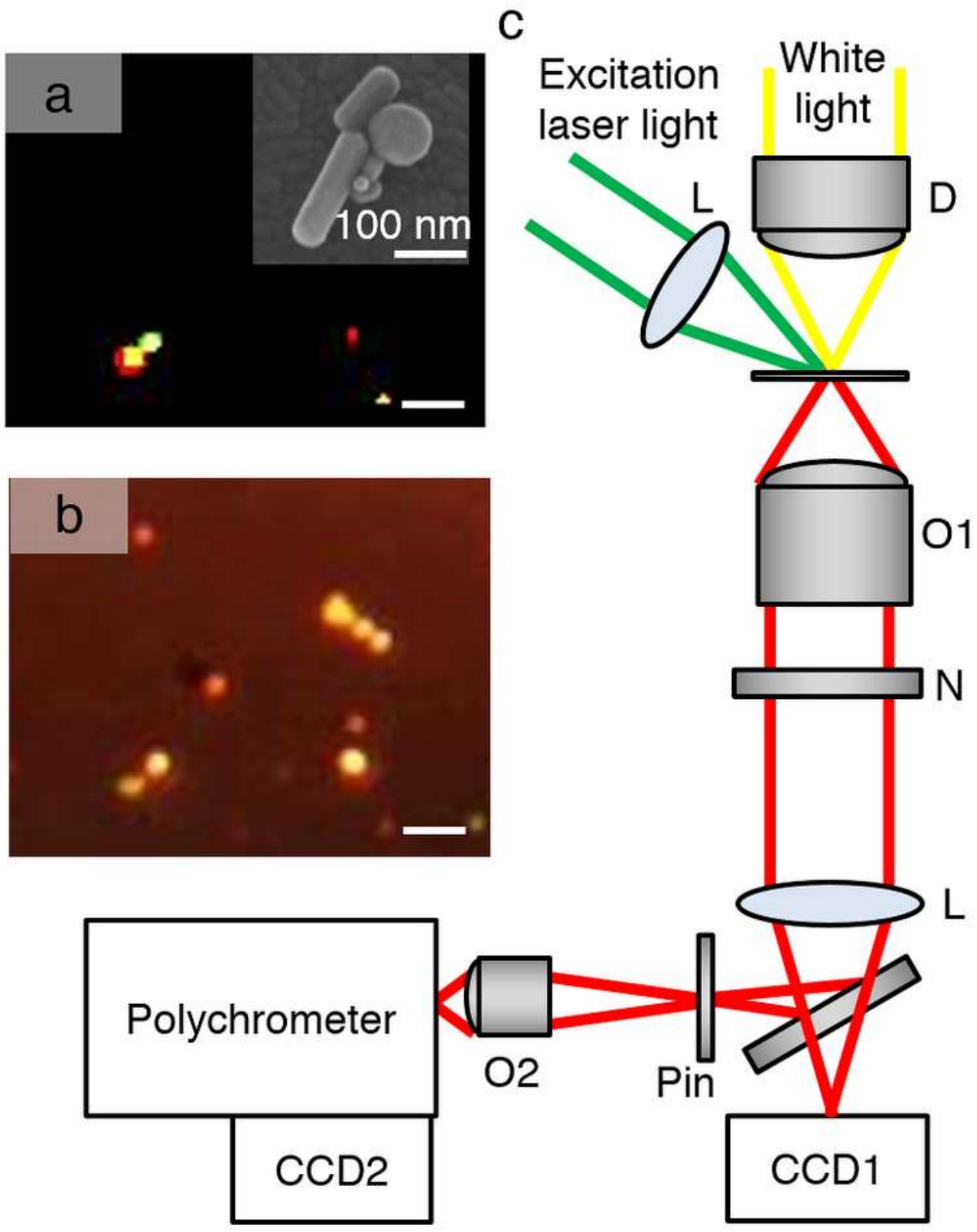

Figure S2. (a) Elastic light scattering image and (b) SERS image of Ag nanoparticle nanoaggregates dispersed on identical ITO glass plate. Scale bars are 5 μm. Inset of (a): a SEM image of the Ag NP nanoaggregates emitting SERS signals. Scale bar is 100 nm. (c) Experimental setup for simultaneous measurement of SERS and elastic light scattering from single Ag NP dimers. A sample aqueous solution of NaCl and Ag NP nanoaggregates adsorbed with 4-ABT were sandwiched between two glass plates. D is a dark-field condenser lens; L is a convex lens; O1 to O3 are objective lenses; N is a notch filter; 'pin' marks a pinhole, and CCD1 and CCD2 are charge-coupled-devices.